\documentclass[
    ,final            
  ]
  {aipproc}

\layoutstyle{8x11double}

\def\gsim{\:\raisebox{-0.5ex}{$\stackrel{\textstyle>}{\sim}$}\:}

\begin{document}

\title{Supersymmetric Higgs singlet effects on FCNC observables}


\author{Robert N. Hodgkinson}{
  address={School of Physics and Astronomy, University of Manchester,
Manchester M13 9PL, United Kingdom}
}

%

\begin{abstract}
Higgs singlet superfields, usually present in extensions of the  Minimal Supersymmetric
Standard Model (MSSM) which address the $\mu$-problem, such as the Next-to-Minimal
Supersymmetric Standard Model (NMSSM) and the Minimal Nonminimal Supersymmetric Standard
Model (mnSSM), can have significant contributions to $B$-meson flavour-changing neutral
current observables for large values of $\tan\beta \gsim 50$.  Illustrative results are
presented including effects on the $B_s$ and on the rare decay $B_s\to\mu^+\mu^-$.  In
particular, we find that in the NMSSM, the branching ratio for $B_s\to\mu^+\mu^-$ can be
enhanced or even suppressed with respect to the Standard  Model prediction by more than
one order of magnitude.
\end{abstract}

\maketitle


\section{Introduction}

The superpotential of the MSSM contains a bilinear term, $\mu \hat H_1 \hat H_2$,
that couples the two Higgs-doublet superfields.  Although successful electroweak
(EW) symmetry breaking implies that $\mu$ must be close to $M_{\rm SUSY}$, $\mu$ can
generally be driven to $M_{\rm GUT}$ or $M_{\rm Planck}$ by supergravity quantum effects.
This theoretical tension is referred to as the $\mu$-problem of the MSSM
\cite{Hall:1983iz,Kim:1983dt}.

We may instead replace the Higgs bilinear term with a trilinear coupling $\lambda
\hat S \hat H_1 \hat H_2$, where $\hat S$ is a new gauge-singlet Higgs superfield.
An effective $\mu$ parameter is generated at the scale $M_{\rm SUSY}$ when $\hat S$
aquires a non-vanishing vacuum expectation value (VEV).
%
Many mechanisms to break the undesirable $U(1)_{\rm PQ}$ symmetry
introduced by this proceedure have been
%
discussed in the literature \cite{Accomando:2006ga} and lead to 
distinct models, such as the Next-to-Minimal Supersymmetric Standard
Model (NMSSM) \cite{Fayet:1974pd,Frere:1983ag,Derendinger:1983bz,Ellis:1988er}
and the Minimal Non-minimal Supersymmetric Standard Model (mnSSM)
\cite{Panagiotakopoulos:2000wp,Dedes:2000jp}.  If $U(1)_{\rm PQ}$ is only weakly
broken then the lightest CP-odd Higgs field $A_1$ of such models becomes a
pseudo-Goldstone boson.  Production of a light
$A_1$ within the context of the NMSSM has been considered through the decays of
SM-like Higgs fields \cite{Ellwanger:2003jt,Chang:2005ht,
Carena:2007jk}, in associated production with charginos \cite{Arhrib:2006sx} and in
rare decays of Upsilon mesons
\cite{
Dermisek:2006py,Fullana:2007uq,Hodgkinson:2008ei}.

It has long been known that 1-loop threshold corrections can produce significant
non-holomorphic Yukawa couplings in the MSSM at large values of $\tan\beta$ \cite{Banks:1987iu,Ma:1988fp,Hempfling:1993kv,Hall:1993gn} and FCNC effects on $B$--meson
observables mediated by MSSM Higgs bosons have been well studied, e.g.
\cite{Babu:1999hn,Buras:2002vd,Dedes:2002er}
Most recently, it has been realized
\cite{Hodgkinson:2006yh,Hodgkinson:2008qk} that analogous threshold corrections produce
sizeable radiative Yukawa couplings for the singlet Higgs bosons in minimal extensions of
the MSSM.  Here we consider the effects of such light singlet Higgs bosons on FCNC
observables within the  MFV framework.

\section{Effective Lagrangian}

The  effective  Lagrangian  describing  the down-type quark  self-energy
transition $Q^0_{jL}\to  d^0_{iR}$ may be written in a gauge-symmetric
and flavour-covariant form as
\begin{eqnarray}
  \label{Formulae:Quark_SE_Lagrangian}
-{\mathcal L}_{\rm eff}^d [\Phi_1,\Phi_2,S]
& = &
\bar{d}^0_{iR}\left({\bf h}_d \Phi_1^\dag
+\Delta{\bf h}_d [\Phi_1,\Phi_2,S]\right)_{ij}
Q^0_{jL}\nonumber\\
& & +\ {\rm H.c.}\ .
\end{eqnarray}
In~(\ref{Formulae:Quark_SE_Lagrangian}) the first term is the tree level
contribution, whilst $\Delta{\bf h}_d$ is a $3\times 3$ matrix which is
a Coleman--Weinberg effective functional
of the
background Higgs fields $\Phi_{1,2}$ and $S$.  The detailed form
of $\Delta{\bf h}_d$ is given in \cite{Hodgkinson:2008qk}.

The interaction Lagrangian may be written in terms of the 
mass eigenstates as
\begin{eqnarray}
 \label{Formulae:Mass_Eigenstate_Couplings_Quarks}
-{\mathcal L}^{d,H}_{\rm FCNC} & = &
\frac{g}{2M_W}\left[H_i \bar d_R
\left(\widehat{\bf M}_d\ {\bf g}^L_{H_i \bar dd}P_L
+{\bf g}^R_{H_i \bar dd}\ \widehat{\bf M}_d P_R\right) d\right.
\nonumber\\
&&\hspace{-0.5cm}
\left.+\ A_j \bar d_R
\left(\widehat{\bf M}_d\ {\bf g}^L_{A_j \bar dd}P_L
+{\bf g}^R_{A_j \bar dd}\ \widehat{\bf M}_d P_R\right) d\right]\; ,
\end{eqnarray}
where the Higgs couplings are given by \cite{Hodgkinson:2008qk}
\begin{eqnarray}
 \label{Formulae:Coupling_Matrices}
{\bf g}^L_{H_i \bar dd} & = &
\frac{{\mathcal O}^H_{1i}}{c_\beta}{\bf V}^\dag
{\bf R}_d^{-1}\left({\bf 1}_3+{\boldmath \Delta}_d^{\phi_1}\right)
{\bf V}
+\frac{{\mathcal O}^H_{2i}}{c_\beta}{\bf V}^\dag
{\bf R}_d^{-1}{\boldmath \Delta}_d^{\phi_2}{\bf V}\nonumber\\
&& +\frac{{\mathcal O}^H_{3i}}{c_\beta}{\bf V}^\dag
{\bf R}_d^{-1}{\boldmath \Delta}_d^{\phi_S}{\bf V}\ ,\\
{\bf g}^L_{A_i \bar dd} & = &
i {\mathcal O}^A_{1i} t_\beta {\bf V}^\dag {\bf R}_d^{-1}
\left({\bf 1}_3+{\boldmath \Delta}_d^{a_1}-\frac{1}{t_\beta}
{\boldmath \Delta}_d^{a_2}\right){\bf V}\nonumber\\
&&-i \frac{{\mathcal O}^{A}_{2i}}{c_\beta}{\bf V}^\dag
{\bf R}_d^{-1}{\boldmath \Delta}_d^{a_S}{\bf V}\ ,
\end{eqnarray}
and $g^R_{H_i(A_i)\bar dd}=\left(g^L_{H_i(A_i)\bar dd}\right)^\dag$.
${\mathcal O}^H$ and ${\mathcal O}^A$ are the scalar and pseudoscalar
mixing matrices respectively.
The resummation matrix ${\bf R}_d$ is defined by
\begin{equation}
 \label{Formulae:Resummation_Matrices}
{\bf R}_d  = {\bf 1}_3\ +\ \frac{\sqrt 2}{v_1}
\left<{\bf h}_d^{-1} \Delta{\bf h}_d\left[\Phi_1,\Phi_2,S\right]
\right>
\ ,\nonumber\\
\end{equation}
where $\left<\ldots\right>$ indicates the VEV of the enclosed expression
and the $3\times 3$ matrices ${\boldmath \Delta}_d^{\phi_{1,2,S}},
{\boldmath \Delta}_d^{a_{1,2,S}}$ are evaluated according to the Higgs
low energy theorem (HLET) \cite{Ellis:2007kb} as
\begin{equation}
 \label{Formulae:HLET}
\frac{{\boldmath \Delta}_d^{\phi_{1,2,S}}}{\sqrt{2}} = 
\left<{\bf h}_d^{-1}\frac{\delta\left(\Delta{\bf h}_d\right)}{\delta\phi_{1,2,S}} \right>,\ 
\frac{{\boldmath \Delta}_d^{a_{1,2,S}}}{\sqrt{2}} = i
\left<{\bf h}_d^{-1}\frac{\delta\left(\Delta{\bf h}_d\right)}{\delta a_{1,2,S}} \right>.
\end{equation}

\vspace{-0.2cm}
\section{Numerical Results and Conclusions}
\setcounter{equation}{0}
\label{NumericalResults}

\vspace{-0.2cm}

We assume the framework of MFV so that all flavour changing effects
are proportional to the CKM matrix ${\bf  V}$.  In calculating the
Higgs couplings, we have used the following benchmark values
\begin{equation}
  \label{Bench}
\begin{array}{l}
{\widetilde M}^2_Q = {\widetilde M}^2_L =  
{\widetilde M}^2_D = {\widetilde M}^2_E = 
{\widetilde M}^2_U\ = (1.7~{\rm TeV})^2,\\
A_u = A_d = A_e = 2.0~{\rm TeV}, \\
M_1 = M_2 = M_3 = 2.0~{\rm TeV}\; , 
\end{array}
\end{equation}
with $\mu=140~{\rm GeV}$ and $t_\beta=50$ throughout.
All points considered are consistent with the $2\sigma$ experimental
bounds on $B\to X_s\gamma$.
%
%

To examine the effects of light singlet Higgs bosons within the mnSSM we take
the parameters describing the tree-level
Higgs sector (with $\mu$ and $\tan\beta$) to be
\begin{equation}
 \label{MNSSMbench}
 M_a=1.5\ {\rm TeV},\quad m_{12}^2=(1.0\ {\rm TeV})^2,\quad 
\lambda=0.3\ .
\end{equation}
The mass scale of the singlet Higgs bosons, $\sim\sqrt{\lambda t_S/\mu}$,
is allowed to vary.
The singlet Higgs scalar and pseudoscalar are approximately degenerate
due to a tree-level mass-sum rule \cite{Panagiotakopoulos:2000wp}.

\begin{figure}[bht!]
\includegraphics[width=0.8\columnwidth]{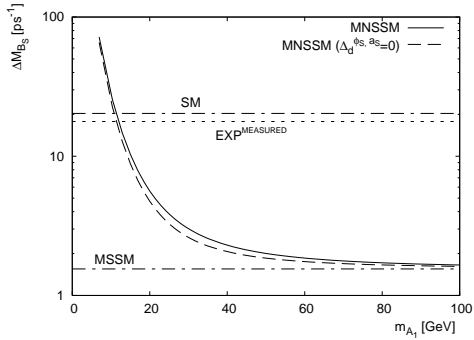}
\caption{\it The SUSY contribution to $\Delta M_{B_s}$ in units of
{\rm ps}$^{-1}$ as a function of the mass of the lightest pseudoscalar
$m_{A_1}$ in the mnSSM. The solid line includes threshold
corrections for the gauge singlet Higgs bosons, the dashed curve
neglects these. All parameters are taken as in~(\ref{Bench})
and~(\ref{MNSSMbench}).
}
\label{MNSSM_MBs}
\end{figure}

In  Fig.~\ref{MNSSM_MBs}  we show  the  SUSY  contribution to  $\Delta
M_{B_s}$
as a  function  of  the  lightest
pseudoscalar  mass.   The upper  curve  includes the  radiative
Yukawa couplings  of the Higgs  singlet fields whilst the  lower curve
neglects these corrections.  The SUSY contribution is seen to 
exceed the currently  observed value if the singlet Higgs bosons are light.
%
%
The dominant contribution to  $\Delta M_{B_s}$ for light singlet Higgs
bosons  is   due  to  the  Wilson   coefficient  $C_1^{\rm  SLL(DP)}$.
Within the mnSSM
there is a cancellation  between the $H_1$-
and  the  $A_1$-mediated  contributions  to $C_1^{\rm  SLL(DP)}$.   An
analogous cancellation between  the heavy Higgs bosons $H$  and $A$ is
known to take  place in the MSSM.
For very light singlets this  cancellation
is dominantly broken by the mass splitting between $H_1$ and $A_1$,
for larger masses this splitting is negligible and the dominant breaking
is due to threshold effects on the singlet-Higgs Yukawa couplings.
\begin{figure}[thb!]
\includegraphics[width=0.8\columnwidth]{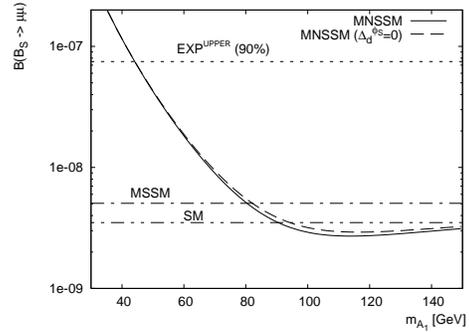}
\caption{\it The branching ratio ${\mathcal B}(B_s\to\mu^+\mu^-)$ as a
function of  the mass  of the lightest  pseudoscalar $m_{A_1}$  in the
mnSSM.  All  parameters  are  taken  as  in~(\ref{Bench})
and~(\ref{MNSSMbench}).
}
\label{MNSSM_Bstomu}
\end{figure}

Figure~\ref{MNSSM_Bstomu} shows the branching ratio ${\mathcal
B}(B_s\to\mu^+\mu^-)$ as a function of $m_{A_1}$.
The prediction exceeds the current bounds for
Higgs singlet masses below around $50$~GeV.  The branching ratio
depends only on the absolute values
of the couplings and so there is no cancellation between the scalar and
pseudoscalar contributions.

Turning to the NMSSM, the CP-odd and CP-even singlets are not in general
constrained to have degenerate masses.  There is typically only one light
Higgs particle, the pseudo-Goldstone boson $A_1$.  There is no cancellation
between the dominant Higgs field contributions to the Wilson coefficients
as in the mnSSM.  This forces us to assume a small singlet-doublet mixing
angle $\cos\theta_A$ at large values of $\tan\beta$.  To examine FCNC
observables within such a scenario, we take $\theta_A$ and $m_{A_1}$
to be free parameters in place of the soft-SUSY breaking trilinears
$A_{\lambda,\kappa}$.  In our resutls we use the following benchmark values;
\begin{equation}
 \label{NMSSMbench}
 \lambda=0.4\,,\quad \kappa=-0.5\,, \quad \cos\theta_A=0.018\; .
\end{equation}

\begin{figure}[thb!]
\includegraphics[width=0.8\columnwidth]{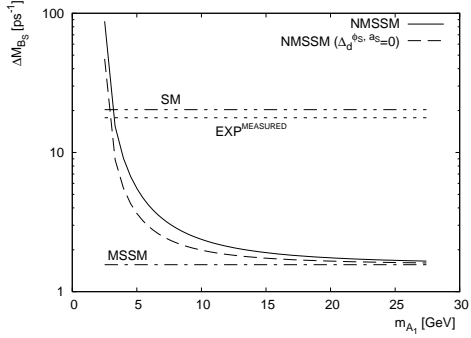}
\caption{\it  The SUSY contribution  to $\Delta  M_{B_s}$ in  units of
{\rm ps}$^{-1}$ as a function of the mass of the lightest pseudoscalar
$m_{A_1}$ in the NMSSM.  The line conventions are as for
Fig.~\ref{MNSSM_MBs}.   All parameters  are taken  as in~(\ref{Bench})
and~(\ref{NMSSMbench}).
}
\label{NMSSM_MBs}
\end{figure}

In Fig.~\ref{NMSSM_MBs} we show the SUSY contribution to
$\Delta M_{B_s}$ as a function of $m_{A_1}$.  In this scenario the
SUSY contribution also exceeds the currently measured value of
$\Delta M_{B_s}$ for the lightest $A_1$ masses.
The small contribution to $\Delta M_{B_s}$  in this scenario
is due to our choice of small $\cos\theta_A$.
As a result of this, the contribution of the singlet pseudoscalar
become negligible  here for $m_{A_1}\gsim  25$~GeV and we  recover the
MSSM prediction.

%

\begin{figure}[thb!]
\includegraphics[width=0.8\columnwidth]{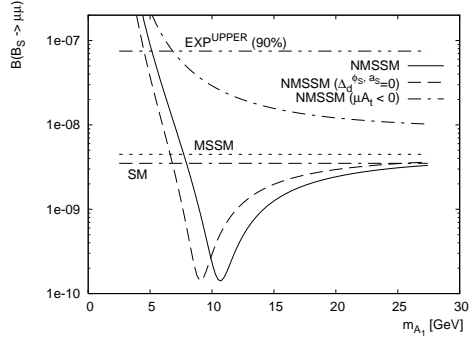}
\caption{\it The branching ratio ${\mathcal B}(B_s\to\mu^+\mu^-)$ as a
function of  the mass of  the lightest pseudoscalar, $m_{A_1}$  in the
NMSSM. The solid line includes threshold corrections for
the gauge singlet Higgs bosons,  the dashed curve neglects these.  All
parameters are taken  as in~(\ref{Bench}) and~(\ref{NMSSMbench}).  The
dot-dashed curve  shows the NMSSM prediction,  including all threshold
corrections,  but with  $A_u=-2$~TeV.
}
\label{NMSSM_Bstomu}
\end{figure}

Figure~\ref{NMSSM_Bstomu} shows the branching ratio ${\mathcal B}
(B_s\to \mu^+\mu^-)$ as a function of the lightest pseudoscalar mass
in the NMSSM. For small values of $m_{A_1}$ the
prediction exceeds the current $90\%$ confidence limits.
For a heavier singlet Higgs pseudoscalar, we observe
that the presence of the singlet leads to a significant reduction
in the branching ratio, which becomes suppressed by more than an
order of magnitude compared to the SM prediction.  Here the contribution
of $A_1$ interferes destructively with the SM-like diagrams.  It
should be noted that such a large suppression is not possible in either
the MSSM or mnSSM, since in both these models tree-level mass-sum
rules prevent the appearance of isolated CP-odd Higgs bosons.
As can be seen from the dot-dashed curve in  Fig.~\ref{NMSSM_Bstomu},
the suppression is  found to vanish when $\mu A_u$ flips sign,
i.e.~for $A_u = -2$~TeV, since the leading SM and SUSY contributions
interfere constructively in this case.

\begin{theacknowledgments}
This  work  was  partially  supported  by  the  STFC  research  grant:
PP/D000157/1.
\end{theacknowledgments}



\bibliographystyle{aipproc}   

\bibliography{neilreferences}

\IfFileExists{\jobname.bbl}{}
 {\typeout{}
  \typeout{******************************************}
  \typeout{** Please run "bibtex \jobname" to optain}
  \typeout{** the bibliography and then re-run LaTeX}
  \typeout{** twice to fix the references!}
  \typeout{******************************************}
  \typeout{}
 }

\end{document}